# Mitigation Plans for the Microbunching-Instability-Related COTR at ASTA/FNAL

A.H. Lumpkin, M. Church, and A.S. Johnson
Mail to: lumpkin@fnal.gov
Fermi National Accelerator Laboratory, Batavia, IL 60510 USA

### 1.1.1 Introduction

At the Advanced Superconducting Test Accelerator (ASTA) now under construction at Fermilab [1], we anticipate the appearance of the microbunching instability related to the longitudinal space charge (LSC) impedances [2,3]. With a photoinjector source and up to two chicane compressors planned, the conditions should result in the shift of some microbunched features into the visible light regime. The presence of longitudinal microstructures (microbunching) in the electron beam or the leading edge spikes can result in strong, spatially localized coherent enhancements of optical transition radiation (COTR) that mask the actual beam profile. Several efforts on mitigation of the effects in the diagnostics task have been identified [4-7]. At ASTA we have designed the beam profiling stations to have mitigation features based on spectral filtering, scintillator choice, and the timing of the trigger to the digital camera's CCD chip. Since the COTR is more intense in the NIR than UV we have selectable bandpass filters centered at 420 nm which also overlap the spectral emissions of the LYSO:Ce scintillators. By delaying the CCD trigger timing of the integration window by 40-50 ns, we can reject the prompt OTR signal and integrate on the delayed scintillator light predominately. This combination of options should allow mitigation of COTR enhancements of order 100-1000 in the distribution.

### 1.1.2 ASTA Facility and Diagnostics Aspects

#### 1.1.2.1 Facility

The base linac planned includes the L-Band photoinjector gun with a $Cs_2Te$ photocathode, two superconducting (SC) rf booster cavities, a chicane, and up to three L-band cryomodules (CM1-3) that each house 8 SCRF cavities. The first cavity of the cryomodule presently installed has been tested to gradients of 31.5 MV/m so one projects a total acceleration capability of 250 MeV per cryomodule. A schematic of the injector is shown in Fig. 1 with a photograph of the shielded tunnel and installed infrastructure in Fig. 2. The gun is driven by the Yb fiber laser oscillator running at 1300 MHz which has been pulse picked down to a 3 MHz micropulse rate, amplified by several single pass amplifiers, and frequency quadrupled to the UV. The macropulse is specified for up to 1 ms length at 5Hz. Charges per micropulse range from 20-3200 pC which are dictated by the UV energy, the quantum efficiency of the cathode, and the experimental requests. At the time of this writing, we are working towards the first testing of the gun with $Cs_2Te$ cathode and the installation of the beamline to the low energy dump. Depending on the status of the first booster cavity we may run first beam tests through only booster cavity 2 which has already been conditioned at about 20 MV/m.

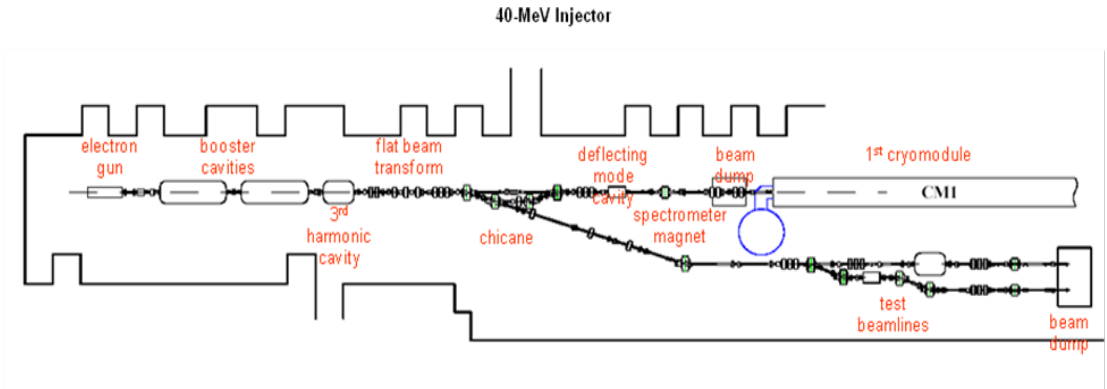

**Figure 1:** A schematic of the ASTA injector showing gun, booster cavities, and chicane for providing beam into CM1. The straight ahead line to the low energy dump is in assembly stage.

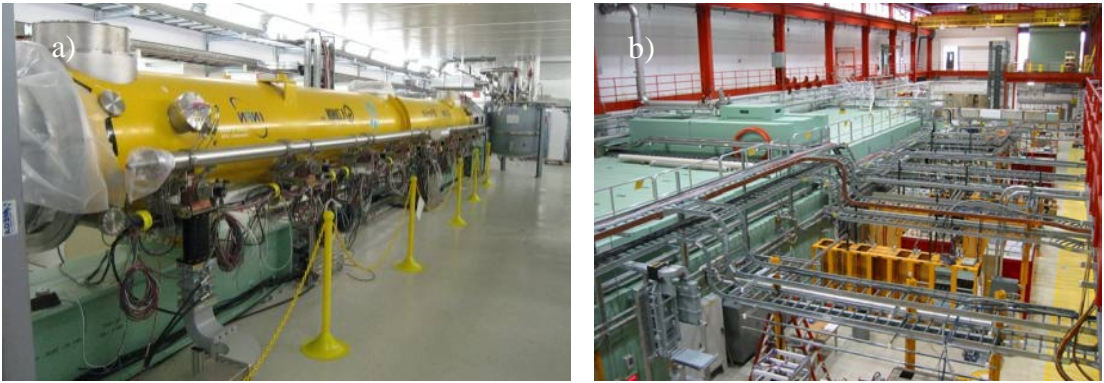

**Figure 2:** Photographs of an installed cryomodule (left) and the shielded tunnel and installed rf and power supply infrastructure (right).

The proposed buildout path in Stage I indicated in Fig. 3 would add the high energy beamline transport to the high power (30 kW) beam dump, install an experimental and diagnostics area, and install the integrable optics test accelerator (IOTA) storage ring. At present the experimental spur beamline at 50 MeV has been postponed.

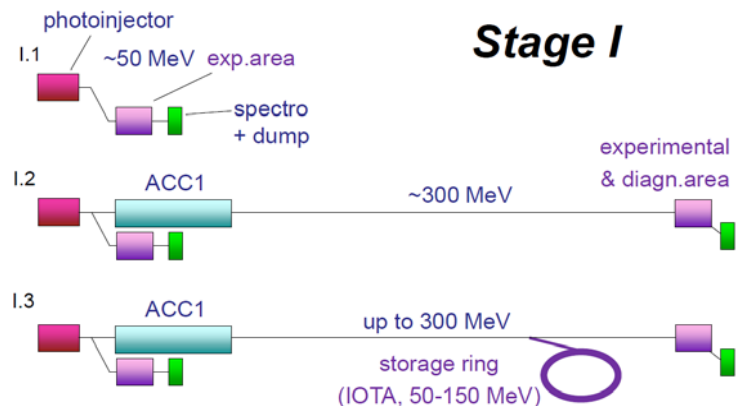

**Figure 3:** Proposed Stage I buildout of the ASTA facility to include the injector linac, high energy experimental area, and the IOTA ring [1].

### *1.1.2.2 Diagnostics Options*

As provided in this workshop's experimental overview talk [8], diagnostics for assessing the µBI via COTR can be developed with standard beam diagnostics with some adaptations.

1) Bunch length monitors for tuning and verifying the compression will be based on coherent radiation aspects of transition radiation (CTR), synchrotron radiation (CSR), edge radiation (CER), diffraction radiation (CDR), etc. in the frequency domain or on incoherent sources in the temporal domain with an ultrafast streak camera or deflecting mode cavity plus an imaging screen. We have planned for a station following the first chicane to provide such capabilities in the injector linac.

2) OTR beam profile monitor screens are used for detecting the presence of COTR and its spatial distribution, intensity fluctuations, and intensity enhancements. The latter can be factors of 100 to 10,000 which make the profiles no longer representative of the true charge distribution and obviate the technique for profiling.

In the event we have COTR, our mitigation techniques include spectral filtering, using the source strength of the scintillator relative to OTR, and temporally sorting the prompt OTR from the delayed scintillator emission with the CCD gate. The spectral aspects are schematically shown in Fig. 4 with the COTR being enhanced in the NIR and the OTR being bluish white to the human eye. A first order mitigation of COTR is provided by a band pass filter centered at 400 nm (violet-rectangle) where the gain is close to one. To improve the signal-to-background ratio, one can employ a scintillator that radiates at within the same filter's transmission bandwidth. Some options are shown in Table 1. We have chosen the Yttrium-doped version of the LSO:Ce crystal, or LYSO:Ce, which also radiates in the 420-nm regime and is commercially available. At ASTA the standard stations after BC1 will have these crystals instead of the YAG:Ce crystals that radiate at 530 nm. We have chosen a scintillator thickness of 100 µm as a trade on efficiency (about 100 times that of OTR) and spatial resolution. Empirical evidence suggests we should have 8-10 µm spatial resolution (sigma) from the scintillator term.

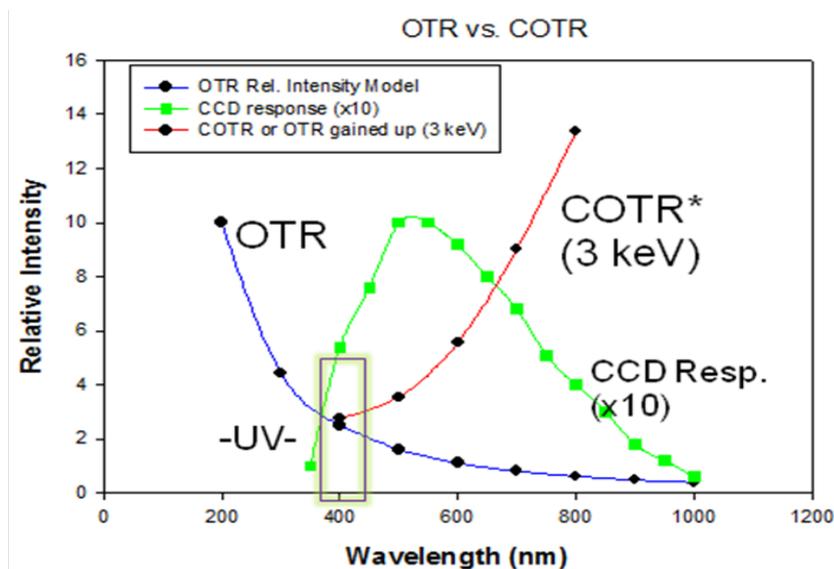

**Figure 4:** Comparison of the OTR and COTR spectral content and the CCD spectral response. The COTR gain is based on a model using the 3-keV slice energy spread [9].

Table 1: Summary of the properties of cerium-doped scintillators as compared to an OTR source.

| Converter | Spectrum (FWHM)*, Peak | Efficiency | Response Time (FWHM) | Comment |
|---|---|---|---|---|
| YAG:Ce | 487-587, 526 nm | 1.0* | 89 ns* | 460 µm T |
| LS0:Ce | 380-450, 415 nm | 0.46* | 40 ns* | 530 µm T |
| YAP:Ce | 350-400, 369 nm | ~0.5 | 28 ns | 460 µm T |
| OTR | Broadband | 0.0013* | ~10 fs | Surface |

The beam profile stations consist of the converter screens on a 4-position pneumatic actuator, the transport optics, and the digital CCD camera as shown in Fig. 5. The positions include an impedance screen, the crystal position, the OTR position, and a calibration target which includes line-pair patterns for checking spatial resolution *in situ.* Our optical imaging resolution term is about 15 µm with an 18-mm FOV. We use the Prosilica 5 Mpix digital cameras with Gig-E format. Image processing is done with a Java-based script online and a Matlab-based script off line. In the production station we use two filter wheels with 5 positions each loaded as listed in Table 2: one to allow the selection of neutral density (ND) filters for signal intensity adjustments and one to select bandpass filters matched to the YAG:Ce or LYSO:Ce scintillator emissions. Additionally, two linear polarizers are available for study of OTR polarization effects and for optimizing the point spread function for the horizontal and vertical planes [10].

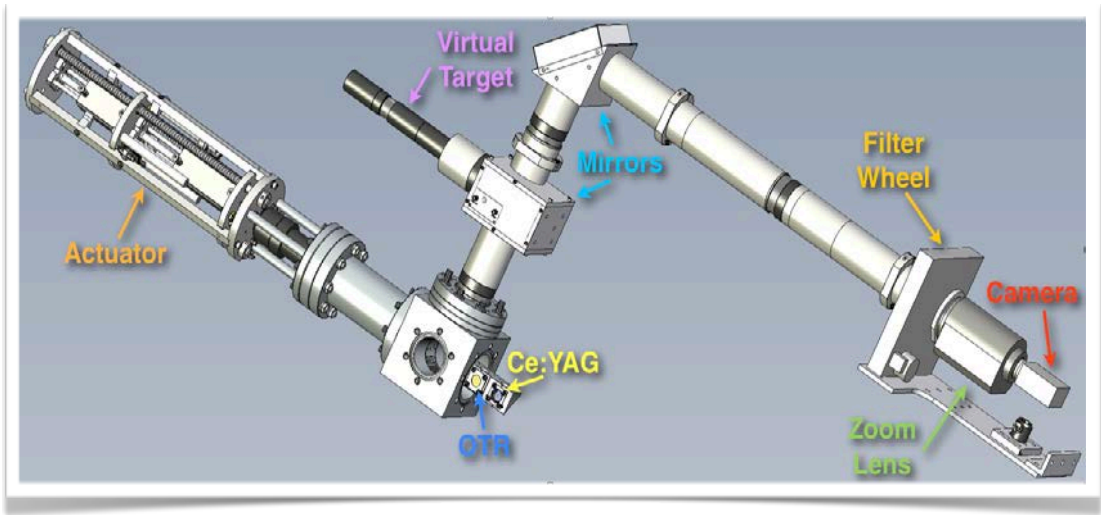

**Figure 5:** The beam profile station prototype showing the vacuum cube, converter screens, optics transport, filter wheel, final lens, and CCD camera.

Table 2: Summary of the options in the two filter wheels at each standard beam profiling station. The uses of filter wheel 2 options are also indicated.

| Position # | Filter Wheel 1 | Filter Wheel 2 | Use |
|---|---|---|---|
| 1 | Clear glass | Clear glass | optics |
| 2 | ND 0.5 | 400x50 nm | LYSO:Ce |
| 3 | ND 1.0 | 550x40 nm | YAG:Ce |
| 4 | ND 2.0 | Horiz. Pol. | OTR |
| 5 | ND 3.0 | Vert. Pol. | OTR |

### 1.1.3 Temporal Mitigation Option in Diagnostics

It has been established previously that one can sort the source terms for radiation with different response times such as the prompt OTR and the delayed emissions of scintillators. Clean separations have been done using the gating feature of the microchannel plate intensifier (MCP) coupled to the CCD camera and single micropulses [6]. Since the MCP may cost over $10k, we pursued a less expensive option based on the digital CCD camera's integration gate. In Fig. 6, the upper images are the LYSO:Ce signal with different trigger delays from the reference time of 742.800 µs. As the trigger moves later in time, the CCD integrated signal level drops as the intensity decays. However for the OTR case, we found a 41 ns delay on the CCD trigger was sufficient to suppress the signal from the prompt OTR as seen in images 6d) and 6e). The rejection ratio is at least 50. For a pulse train in the linac, a fast pulse kicker could be used to direct a single micropulse to the off-axis imaging station.

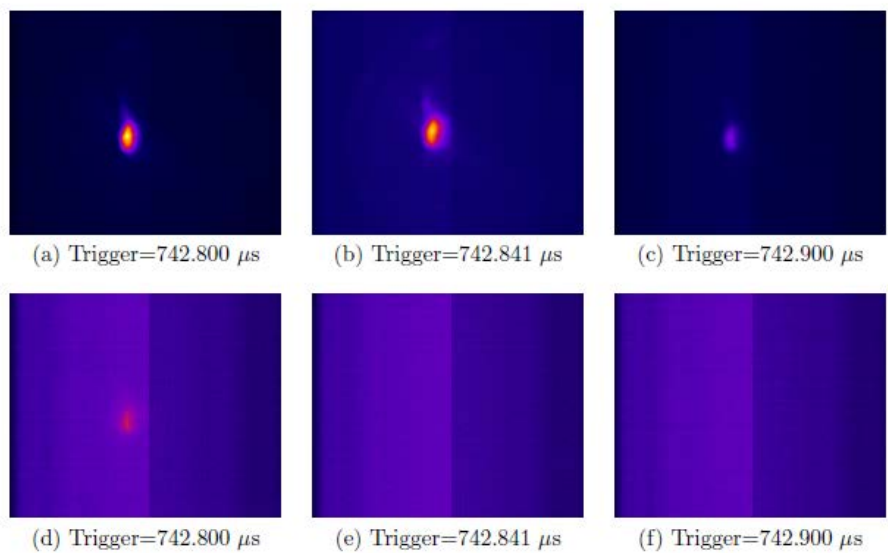

**Figure 6:** Beam images for different CCD trigger times for the chip integration period for the LYSO:Ce scintillator in (a-c) and the OTR source (d-f).

### 1.1.4 Summary

In summary, based on the experiences at other laboratories [8], we anticipate the µBI will be present in our photo-injected beams at ASTA. At a minimum, the appearance of COTR due to the instability is expected after the second compressor. We also have the option to track the visibility of the COTR over a range of charges from 20 to 3200 pC per micropulse. We plan to use the spectral differences between OTR and COTR, the scintillators in combination with bandpass filters to enhance the signal-to-background ratio, and temporal gating techniques to mitigate the diagnostics effects by a total factor of 100-1000. This should address this diagnostics issue sufficiently to provide reliable beam profile measurements under such conditions.

### 1.1.5 Acknowledgments

The authors acknowledge the support of N. Eddy and R. Dixon of Fermilab and the production of the beam profiling stations by RadiaBeam based on Fermilab specifications.